\pdfoutput=1
\documentclass[12pt]{iopart}

\usepackage{graphicx}
\usepackage{iopams}  

\newcommand{\elab}{E_{lab} = 35\mathrm{~A~GeV}}

\newcommand{\stsqn}{\sqrt{s_{\mathrm{NN}}} = 7.7\mathrm{~GeV}}

\newcommand{\ptt}{p_\mathrm{T}}

\begin{document}

\title[Elliptic flow of inclusive charged hadrons]{Elliptic flow of inclusive charged hadrons in Au+Au collisions at $\elab$ using the PHSD model}

\author{Waseem Bhat$^1$, M. Farooq Mir$^1$, Vipul Bairathi$^2$, Towseef Bhat$^1$, Sonia Kabana$^2$, and Shabir Bhat$^1$}
\address{$^1$ University of Kashmir, Srinagar 190006, India}
\address{$^2$ Instituto de Alta Investigación, Universidad de Tarapacá, Casilla 7D Arica 1000000, Chile}
\ead{vipul.bairathi@gmail.com}
\vspace{10pt}

\begin{indented}
\item[]February 2023
\end{indented}

\begin{abstract}
Elliptic flow ($v_2$) measurements of inclusive charged hadrons at mid-rapidity ($|\eta| <$ 1.0) in Au+Au collisions at $\elab$ using the Parton Hadron String Dynamics (PHSD) model are presented as a function of centrality, transverse momentum ($\ptt$) and pseudo-rapidity ($\eta$). The $v_2$ results are obtained using the $\eta$-sub event plane method with respect to event plane angle ($\psi_{2}$) and participant plane angle ($\psi_{2}^{PP}$). $\ptt$-integrated charged hadron $v_2$ shows a strong centrality dependence in Au+Au collisions at $\elab$. The eccentricity scaled elliptic flow ($v_2/\varepsilon_2$) also shows centrality dependence. The higher values of $v_2/\varepsilon_2$ in central collisions suggest development of stronger collectivity. The measurements are compared with the results from Au+Au collisions at $\stsqn$ published by the STAR experiment at RHIC. We also compare results of HSD and PHSD modes of the model to investigate contribution of hadronic and partonic phases of the medium on the measured $v_2$. The current results serve as a prediction of the collective behavior of the matter produced in baryon rich and moderate temperature conditions for the upcoming Compressed Baryonic Matter (CBM) experiment at the Facility for Antiproton and Ion Research (FAIR). These predictions are also useful for the interpretation of data measured at RHIC Beam Energy Scan (BES) program.
\end{abstract}

\vspace{2pc}

\section{Introduction}
\label{sec:Intro}
Lattice quantum chromodynamics (lQCD) predicts a phase transition from normal nuclear matter to a novel state of matter composed of de-confined quarks and gluons, called the quark gluon plasma (QGP)~\cite{ref1a,ref1b,ref1c}. The experimental facilities like Relativistic Heavy Ion Collider (RHIC)~\cite{ref2a,ref2b,ref2c,ref2d} and Large Hadron Collider (LHC)~\cite{ref2e,ref2f,ref2g} are designed to collide heavy-ions at high temperatures and low net baryon densities to study properties of the QGP. The upcoming CBM experiment at FAIR aims to operate at moderate temperatures and high net baryon densities for studying the QGP medium created in heavy-ion collisions. It will collide Au-ions at beam energies from 2 A GeV to 35 A GeV~\cite{ref3}. In these collisions, the densities are expected to reach 6 to 12 times the ordinary nuclear matter density at the point of collision, which could result in formation of the QGP medium~\cite{ref4,ref5}. 

Various observables have been studied experimentally as well as theoretically to probe the QGP medium created in relativistic heavy-ion collisions. Collective flow is one such observable, which plays a vital role in understanding properties of the QGP medium~\cite{ref6,ref7}. In non-central nucleus-nucleus collisions, the initial spatial anisotropy transforms into momentum space anisotropy due to multi-particle interactions among the constituents of the medium. The momentum space anisotropy can be measured by azimuthal angle distribution of the produced particles with respect to the reaction plane. The reaction plane is the plane formed by the impact parameter vector and the beam direction. The azimuthal angle distribution of the produced particles can be expanded in terms of a Fourier series as, 
\begin{equation}
\frac{dN}{d(\phi-\Psi_{R})} \propto 1 + 2 \sum_{n=1}^{\infty} v_{n} \cos\left[n\left(\phi - \Psi_{R}\right)\right]. 
\label{eq:1}
\end{equation}

The $2^{nd}$-order Fourier coefficient $v_2$, known as elliptic flow, provides a strong evidence for the formation of QGP matter~\cite{ref6,ref8,ref9,ref10}. The charged hadron $v_2$ measurements suggest a hydrodynamic behavior of the QGP with a very low shear viscosity to entropy density ratio ($\eta/s$)~\cite{ref11,ref12}. Further, it is also sensitive to the early phases of the collision and level of thermalization achieved by the system created in heavy-ion collisions~\cite{ref10,ref11,ref14a,ref14b}. Therefore, the charged hadron $v_{2}$ as a function of centrality, transverse momentum ($p_{T}$) and pseudo-rapidity ($\eta$) can provide information about the medium created in heavy-ion collisions.

In this paper, we report measurements of inclusive charged hadron elliptic flow at mid-rapidity ($|\eta| < 1.0$) in Au+Au collisions at $E_{lab}$ = 35 A GeV ($\sqrt{s_{NN}} \approx$ 8.0 GeV) using the PHSD model~\cite{ref18,ref19,ref20}. These $v_{2}$ measurements are the first predictions for the FAIR energy ($E_{lab}$ = 35 A GeV) using the PHSD model. The $v_{2}$ results are compared with the published STAR experimental data from Au+Au collisions at $\sqrt{s_{NN}}$ = 7.7 GeV~\cite{ICh}.  We discuss centrality dependence of $p_{T}$-integrated elliptic flow ($\langle v_2 \rangle$). The differential elliptic flow as a function of $p_{T}$ and $\eta$ are studied for different centrality classes. We also study eccentricity scaled elliptic flow ($v_{2}/\epsilon_{2}$) and discuss collectivity in Au+Au collisions at $E_{lab}$ = 35 A GeV. We study effect of partonic and hadronic interactions by comparing the $v_{2}$ results from partonic (PHSD) and hadronic (HSD) modes of the PHSD model.

The paper is organized in the following way. In section~\ref{sec:model}, we describe the PHSD model in brief. The analysis method is presented in section~\ref{sec:analysis}.
In section~\ref{sec:results}, we present the results on integrated and differential $v_2$. Eccentricity scaled elliptic flow ($v_{2}$/$\varepsilon_{2}$) is discussed. A comparison of $v_2$($p_{T}$) between the HSD and PHSD modes of the PHSD model is also investigated. Finally, we summarize and conclude the results in section~\ref{sec:conclusions}.

\section{PHSD model}
\label{sec:model}
PHSD model is a microscopic covariant transport method that describes highly interacting hadronic and partonic matter produced in heavy-ion collisions~\cite{ref18,ref19,ref20}. It is a consistent dynamical approach formulated on the basis of Kadanoff-Baym (KB) equations~\cite{ref21} or off-shell transport equations in phase-space representation. The Dynamical Quasi-Particle Model (DQPM)~\cite{ref23,ref24,ref25} is designed to describe the strongly interacting non-perturbative nature of the QCD matter. The DQPM is mapped to replicate lQCD results~\cite{ref26,ref27} for a quark-gluon plasma in thermodynamic equilibrium and  serves as the foundation for the transport of theoretical description of the QGP phase in the PHSD model. The DQPM is thermodynamically consistent (contradictory to the massless pQCD partons) in which strongly interacting quarks and gluons represent the degrees of freedom. The hadronization procedure complies with all applicable conservation laws (flavor current and four-momentum) because of off-shell nature of hadrons and partons  in each event~\cite{ref28}. The off-shell Hadron-String Dynamics (HSD)\cite{ref29,ref30} dynamics, which incorporates self-energies for the hadronic degrees of freedom, governs the hadronic system. The low-energy hadron-hadron collisions are modeled in accordance with experimental cross sections, while as inelastic hadron-hadron collisions with energies exceeding $\sqrt{s_{NN}} \geq$ 2.6 GeV are represented by using FRITIOF 7.02 model~\cite{ref31} and PYTHIA 6.4~\cite{ref32}.

\begin{figure}[!htbp]
\begin{center}
\includegraphics[width=9cm]{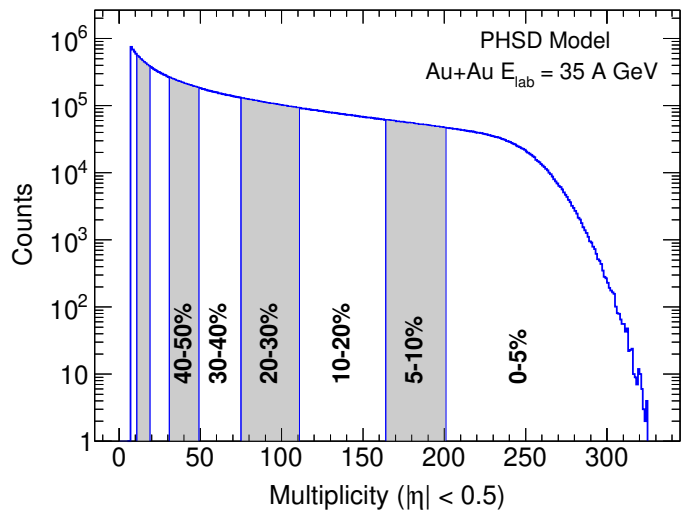}
\caption{(Color online) Reference multiplicity distribution in Au+Au collisions at $E_{lab}$ = 35 A GeV using the PHSD model.}
\label{fig:refmult}
\end{center}
\end{figure}

We have simulated 50 million Au+Au collision events at $E_{lab}$ = 35 A GeV using the PHSD model. We have employed 4.1 version of the PHSD model which incorporates both partonic (PHSD) and hadronic (HSD) modes. Impact parameter range from 0 to 15 $fm$ has been used to generate the events. We have taken the hadronic cascade time of 500 $fm/c$ for the elliptic flow calculation. All the measurements are done in the mid-rapidity ($|\eta| < 1.0$) region. In this study, charged particle multiplicity within $|\eta|<$ 0.5 is selected for the determination of centrality. The reference multiplicity distribution is shown in figure~\ref{fig:refmult} for Au+Au collisions at $E_{lab}$ = 35 A GeV using the PHSD model. The multiplicity is divided into nine different centrality classes from $0-80\%$.

\section{Flow Analysis Method}
\label{sec:analysis}
\subsection{Event Plane Method}
The event plane method is generally used for the measurement of elliptic flow ($v_{2}$) in heavy-ion collisions~\cite{refEPM,refEPM1}. In this method, $v_{2}$ is calculated with respect to $2^{nd}$ harmonic event plane angle $\psi_{2}$ which is given by,
\begin{equation}
\psi_{2} = \frac{1}{2} \tan^{-1}\frac{\sum_{i}w_{i}sin(2\phi_{i})}{\sum_{i}w_{i}     \cos(2\phi_{i})},
\label{eq:2}
\end{equation}

where, $\phi_{i}$ and $w_{i}$ represent azimuthal angle and weight for the $i^{th}$ particle, respectively. We have taken transverse momentum as the weight to optimize the event plane resolution. After the reconstruction of the event plane angle, the $v_{2}$ is calculated using the equation,
\begin{equation}
v_{2} = \left\langle \cos\left[2(\phi - \psi_{2}\right)] \right\rangle.
\end{equation}

The angle brackets denote an average over all the produced particles in all events. Due to limited number of particles used in the calculation of the event plane angle, the estimated event plane is different than the reaction plane. In order to account for this effect, the observed $v_{2}$ is divided by the event plane resolution. For eliminating the auto-correlation and non-flow effects, the $\eta$-sub event plane method is used~\cite{refEPM}. In this method, each event is split into two equal multiplicity sub events with an $\eta$ gap of $0.15$ between them. For each event, the sub-event plane angles are calculated in the positive ($1.0 < \eta < 0.075$) and negative ($-0.075 < \eta < -1.0$) pseudo-rapidity regions. Additionally, we have also calculated $v_{2}$ with respect to the participant plane angle ($\psi_{2}^{PP}$) and $\psi_{2} = 0$.

\subsection{Event plane resolution}
\begin{figure}[!htbp]
\begin{center}
\includegraphics[width=9cm]{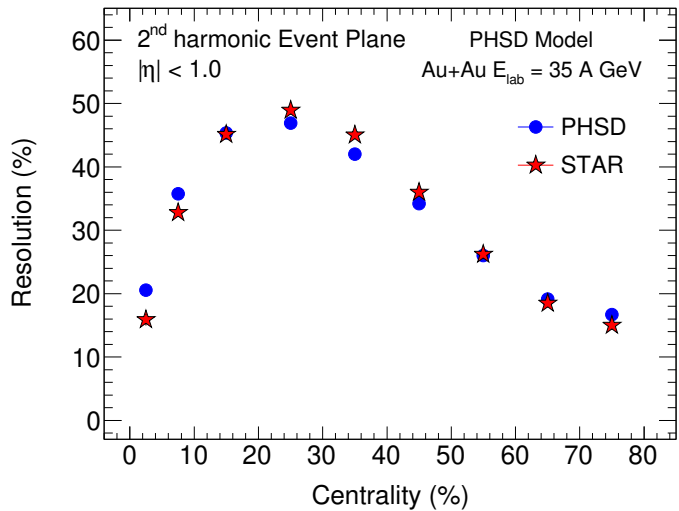}
\caption{(Color online) Event plane angle resolution as a function of centrality at mid-rapidity ($|\eta| < 1.0$) in Au+Au collisions at $E_{lab}$ = 35 A GeV using the PHSD model. The resolution from the STAR experiment in Au+Au collisions at $\sqrt{s_{NN}}$ = 7.7 GeV is also shown~\cite{ICh}.}
\label{fig:epres}
\end{center}
\end{figure}

The event plane angle resolution as a function of centrality at mid-rapidity ($|\eta| < 1.0$) in Au+Au collisions at $E_{lab}$ = 35 A GeV using the PHSD model is shown in figure~\ref{fig:epres}. For comparison with the STAR experimental data, we have converted the sub event plane angle resolution into full event plane angle resolution ($R_{Full} = \sqrt{2} \times R_{\eta-sub}$), where $R_{Full}$ is the full event plane resolution and $R_{\eta-sub}$ represents the eta-sub event plane resolution. The event plane resolution from the PHSD model are matching well with the experimental data from Au+Au collisions at $\sqrt{s_{NN}}$ = 7.7 GeV for all centralities except for the most central collisions, where a small deviation is observed. The resolution peaks at the 20-30\% centrality class and decreases towards the central and peripheral collisions. This is due to the two competing effects of low number of produced particles in peripheral collisions and small $v_{2}$ values in central collisions.

\section{Results}
\label{sec:results}
\subsection{Inclusive charged hadron elliptic flow}
\begin{figure}[!htbp]
\begin{center}
\includegraphics[width=9cm]{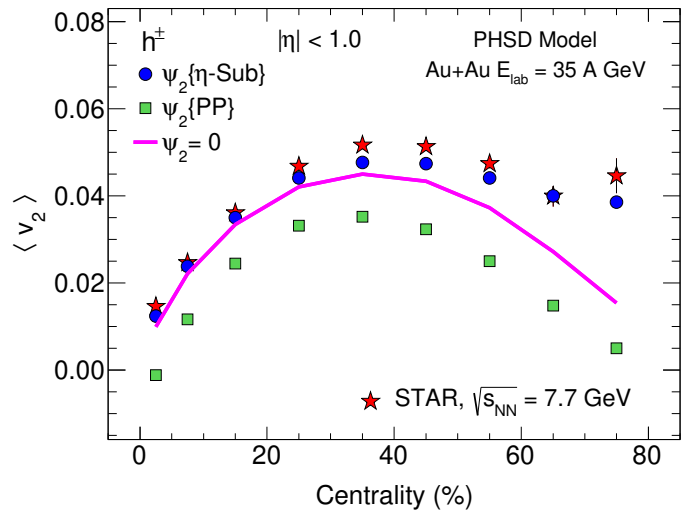}
\caption{(Color online) Integrated elliptic flow ($\langle v_{2} \rangle$) for inclusive charged hadrons as a function of centrality at mid-rapidity ($|\eta| < 1.0$) in Au+Au collisions at $E_{lab}$ = 35 A GeV using the PHSD model. The results for the inclusive charged hadron $\langle v_{2} \rangle$ from the STAR experimental data in Au+Au collisions at $\sqrt{s_{NN}}$ = 7.7 GeV is also shown~\cite{ICh}.}
\label{fig:avgv2}
\end{center}
\end{figure}

Integrated elliptic flow $\langle v_{2} \rangle$ for inclusive charged hadrons as a function of centrality at mid-rapidity ($|\eta| < 1.0$) in Au+Au collisions at $E_{lab}$ = 35 A GeV using the PHSD model is shown in figure~\ref{fig:avgv2}. The obtained $\langle v_{2} \rangle$ shows a clear centrality dependence where the values increase from central to mid-central collisions and then decrease slowly for peripheral collisions. The maximum value of $\langle v_{2} \rangle$ is around 30-40\% centrality. The $\langle v_{2} \rangle$ results are compared with the published STAR results from Au+Au collisions at $\sqrt{s_{NN}}$ = 7.7 GeV~\cite{ICh}. We observed the $\langle v_{2} \rangle$ values follow a similar centrality dependence and are in good agreement with the published STAR experimental data within the statistical uncertainties. 

We compare $\langle v_{2} \rangle$ of charged hadrons calculated with respect to $\psi_{2}$ = 0 in Au+Au collisions at $E_{lab}$ = 35 A GeV using the PHSD model as shown in figure~\ref{fig:avgv2}. The values of $\langle v_{2} \rangle$ for $\psi_{2}$ = 0 are similar to $\langle v_{2} \rangle$ values calculated with respect to $\psi_{2}$ in central collisions (0-20\%) but there are differences in mid-central and peripheral collisions. The difference further increases from mid-central to peripheral collisions, which could be due to the assumption of $\psi_{2}$ = 0 for each event. We also compare $\langle v_{2} \rangle$ calculated with respect to the participant plane angle ($\psi_{2}^{PP}$) as shown in the figure~\ref{fig:avgv2}. We observed the magnitude of $\langle v_{2} \rangle$ with respect to $\psi_{2}^{PP}$ is lower than the $\langle v_{2} \rangle$ with respect to $\psi_{2}$ for all the centralities. This difference might be due to event-by-event fluctuations in the positions of nucleons used for the calculation of participant plane angle. 

\begin{figure}[!htbp]
\begin{center}
\includegraphics[width=12cm]{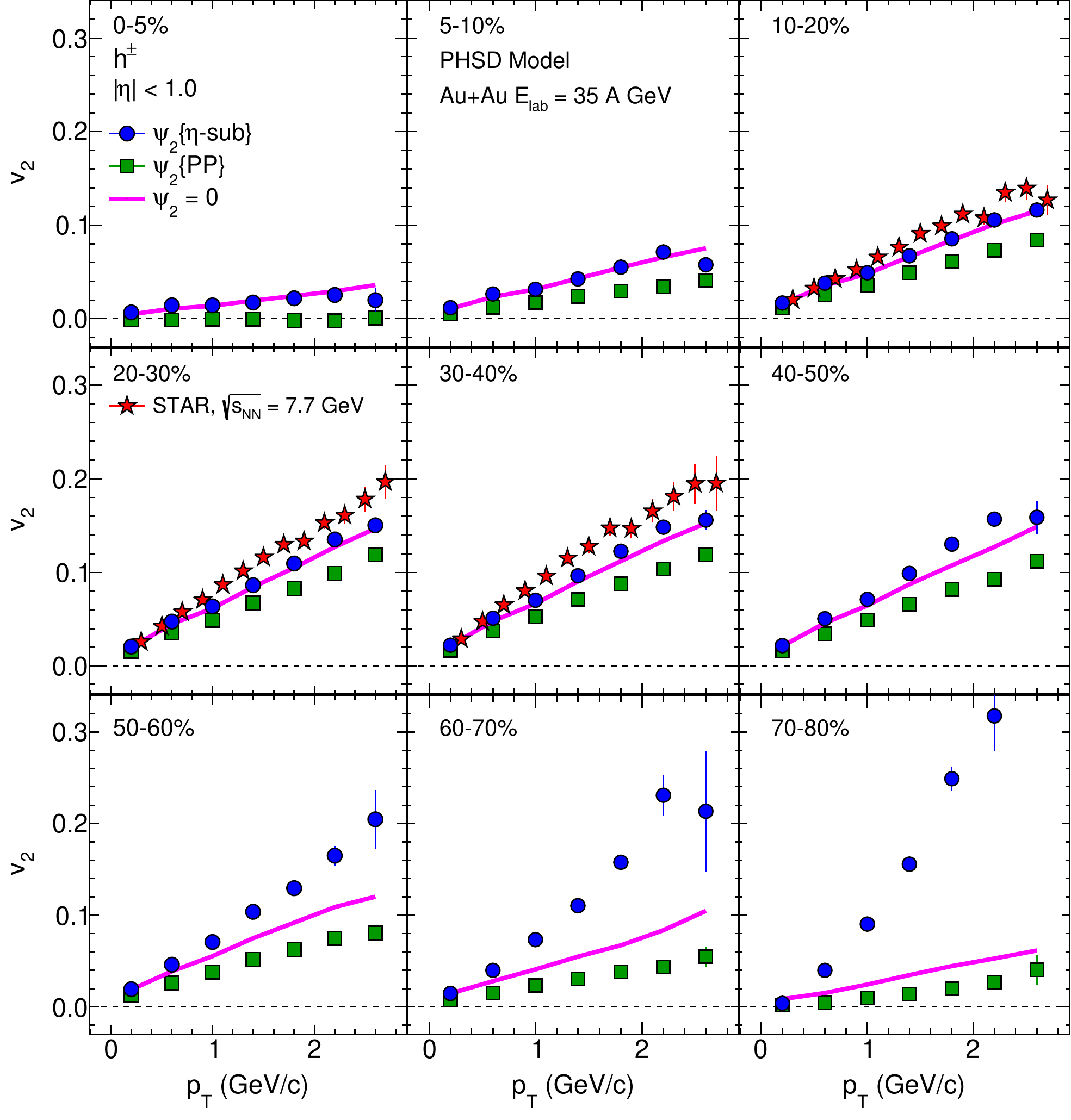}
\caption{(Color online) Differential elliptic flow ($v_{2}$) as a function of $p_{T}$ for charged hadrons at mid-rapidity ($|\eta| < 1.0$) in Au+Au collisions at $E_{lab}$ = 35 A GeV using the PHSD model. Charged hadron $v_{2}$($p_{T}$) from the STAR experimental data in Au+Au collisions at $\sqrt{s_{NN}}$ = 7.7 GeV is also shown~\cite{ICh}.}
\label{fig:v2pt}
\end{center}
\end{figure}

We also report transverse momentum ($p_{T}$) dependence of inclusive charged hadron $v_{2}$ at mid-rapidity ($|\eta| < 1.0$) in Au+Au collisions at $E_{lab}$ = 35 A GeV using the PHSD model as shown in figure~\ref{fig:v2pt}. There is a monotonic increase in $v_{2}$ with $p_{T}$ for all centrality classes. We observed a centrality dependence of $v_{2}$($p_{T}$), where the values increase from central to peripheral collisions for a given $p_{T}$. The measured $v_{2}$($p_{T}$) is compared with the experimental data from Au+Au collisions at $\sqrt{s_{NN}}$ = 7.7 GeV for three centrality classes 10-20\%, 20-30\%, and 30-40\%~\cite{ICh}. We observed the $v_{2}$($p_{T}$) agrees well with the experimental data within the statistical uncertainties for all the three centrality classes. Additionally, we discuss $v_{2}$($p_{T}$) measured with respect to $\psi_{2}^{PP}$ and $\psi_{2} = 0$. We observed that $v_{2}$($p_{T}$) increases monotonically for all the centrality classes. However, the magnitude of $v_{2}$($p_{T}$) calculated with respect to $\psi_{2}^{PP}$ and $\psi_{2} = 0$ increases with $p_{T}$ from central to mid-central collisions and then decreases towards peripheral collisions.

\subsection{Eccentricity scaling}
\begin{figure}[!htbp]
\begin{center}
\includegraphics[width=9cm]{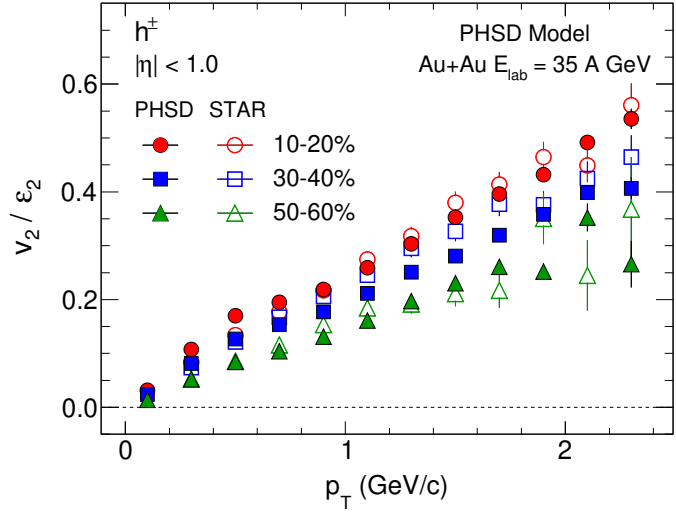}
\caption{(Color online) $v_{2}/\varepsilon_{2}$ as a function of $p_{T}$ for charged hadrons at mid-rapidity ($|\eta| < 1.0$) in Au+Au collisions at $E_{lab}$ = 35 A GeV using the PHSD model. $v_{2}/\varepsilon_{2}$ for charged hadrons from the STAR experiment in Au+Au collisions at $\sqrt{s_{NN}}$ = 7.7 GeV is also shown~\cite{ICh}.}
\label{fig:v2e2}
\end{center}
\end{figure}

In this section, we discuss eccentricity-scaled elliptic flow ($v_{2}/\varepsilon_{2}$) of charged hadrons as a function of $p_{T}$ in Au+Au collisions at $E_{lab}$ = 35 A GeV using the PHSD model as shown in figure~\ref{fig:v2e2}. The obtained $v_{2}$ is divided by the average participant eccentricity ($\varepsilon_{2}$) in each centrality bin. This scaling will remove the contribution of geometry of the initial overlap region. We observe $v_{2}/\varepsilon_{2}$ increases monotonically with increasing $p_{T}$ for all centrality classes. For a given $p_{T}$, the value of $v_{2}/\varepsilon_{2}$ is more in central collisions (10-20\%) compared to peripheral collisions (50-60\%). This observation suggests a stronger collectivity in central collisions. The $v_{2}/\varepsilon_{2}$ is compared with the STAR results from Au+Au collisions at $\sqrt{s_{NN}}$ = 7.7 GeV~\cite{ICh}. The results are consistent with the STAR experimental data within the statistical uncertainties for the measured $p_{T}$ range.

\subsection{Pseudo-rapidity dependence of $v_2$}
\begin{figure}[!htbp]
\begin{center}
\includegraphics[width=9cm]{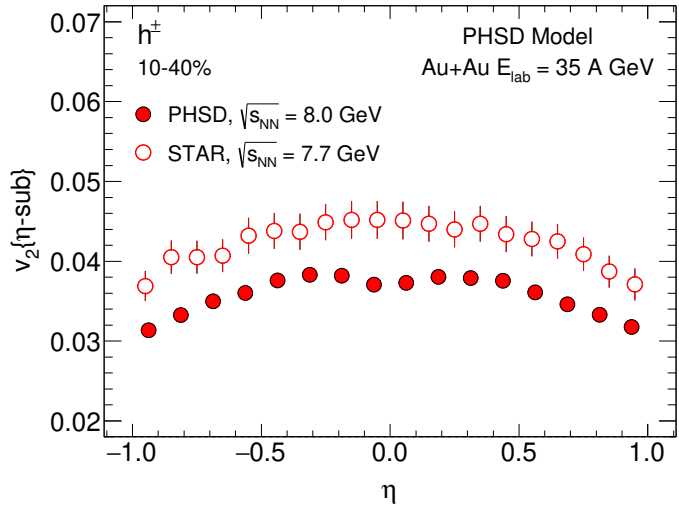}
\caption{(Color online) Inclusive charged hadron $v_{2}$ as a function of $\eta$ in 10-40\% central Au+Au collisions at $E_{lab}$ = 35 A GeV using the PHSD model. Charged hadron $v_{2}$($\eta$) from the STAR experiment in 10-40\% central Au+Au collisions at $\sqrt{s_{NN}}$ = 7.7 GeV is also shown~\cite{ICh}.}
\label{fig:v2eta}
\end{center}
\end{figure}

The inclusive charged hadron $v_{2}$ as a function of pseudo-rapidity ($\eta$) is measured in 10-40\% central Au+Au collisions at $E_{lab}$ = 35 A GeV using the PHSD model as shown in figure~\ref{fig:v2eta}. The $v_{2}$ values are symmetric about $\eta \sim 0$ in the measured pseudo-rapidity range ($-1.0 < \eta < 1.0$). The $v_{2}$ does not change significantly over the measured $\eta$ region. The $v_{2}$($\eta$) results from the PHSD model are compared with the STAR experimental data from 10-40\% central Au+Au collisions at $\sqrt{s_{NN}}$ = 7.7 GeV~\cite{ICh}. We observe the values of $v_{2}$ follow a similar $\eta$ dependence as in the experimental data. However, the magnitude of $v_{2}$($\eta$) in Au+Au collisions at $E_{lab}$ = 35 A GeV from the PHSD model is lower compared to the STAR experimental data.

\subsection{Mode comparison}
\begin{figure}[!htbp]
\begin{center}
\includegraphics[width=12cm]{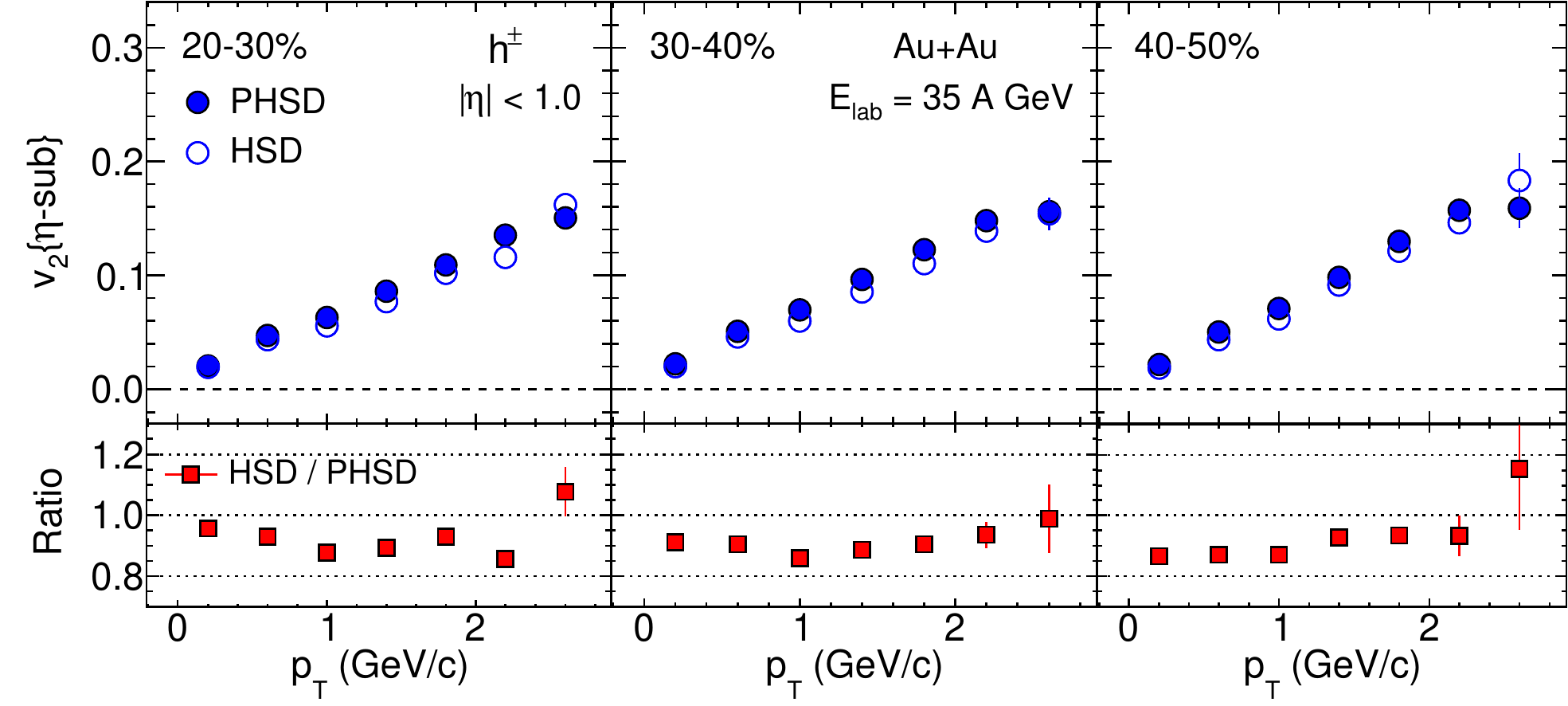}
\caption{(Color online) Inclusive charged hadron $v_{2}$($p_{T}$) at mid-rapidity ($|\eta| < 1.0$) in Au+Au collisions from the HSD and PHSD modes of the PHSD model at $E_{lab}$ = 35 A GeV. The bottom panels show the ratio of $v_{2}$($p_{T}$) between the HSD and PHSD modes.}
\label{fig:v2mod}
\end{center}
\end{figure}

In this sub-section, we compare $v_{2}$($p_{T}$) between the hadronic (HSD) and partonic (PHSD) modes of the PHSD model. Figure~\ref{fig:v2mod} shows the inclusive charged hadron $v_{2}$ as a function of $p_{T}$ at mid-rapidity ($|\eta| < 1.0$) in Au+Au collisions at $E_{lab}$ = 35 A GeV. The HSD mode incorporates only the hadronic interactions, while the PHSD mode incorporates both hadronic as well as partonic interactions. We calculate the ratio of $v_{2}$($p_{T}$) between the HSD and PHSD modes as shown in lower panels of figure~\ref{fig:v2mod}. The $v_{2}$($p_{T}$) increases with $p_{T}$ in both the modes for all centralities. However, the ratio between the two modes is less than unity which shows $v_{2}$($p_{T}$) from the HSD mode is small as compared to the PHSD mode. This difference could be attributed to the partonic interactions in addition to the hadronic interactions in the PHSD mode. This observation indicates the formation of QGP in Au+Au collisions at $E_{lab}$ = 35 A GeV in the PHSD model.

\section{Summary and conclusions}
\label{sec:conclusions}
We have reported elliptic flow measurements of inclusive charged hadrons at mid-rapidity in Au+Au collisions at $E_{lab}$ = 35 A GeV using the PHSD model. The obtained $v_{2}$ is calculated for nine centrality intervals from 0 to 80\% using the $\eta$-sub event plane method. These $v_{2}$ measurements of inclusive charged hadrons are the first predictions in Au+Au collisions at FAIR energy ($E_{lab}$ = 35 A GeV) using the PHSD model. The integrated charged hadron $\langle v_{2} \rangle$ shows a clear centrality dependence and is consistent with the published STAR experimental data~\cite{ICh}. The differential elliptic flow increases with $p_{T}$ for all the centrality classes studied. The observed $p_{T}$ dependence of $v_{2}$ for 10-20\%, 20-30\%, and 30-40\% centrality classes is found to agree well with the STAR experimental results within the statistical uncertainties~\cite{ICh}. The magnitude of eccentricity scaled elliptic flow ($v_{2}/\varepsilon_{2}$) is more in central collisions than mid-central and peripheral collisions which suggests a stronger collectivity in central collisions. Charged hadron $v_{2}$ as function of $\eta$ shows weak dependence in the measured $\eta$ range. However, there is a significant difference between the obtained $v_{2}$($\eta$) calculated from the PHSD model and the published STAR experimental data~\cite{ICh}. In addition, a comparison of $v_{2}$($p_{T}$) between the HSD and PHSD modes show that the magnitude of $v_{2}$($p_{T}$) is larger in the PHSD mode compared to HSD mode. This observation suggests the formation of QGP in the initial stages of the collision at $E_{lab}$ = 35 A GeV. The collision energy in this study has been selected to match the energy that will be available in the future CBM experiment at FAIR. Our results would be helpful to predict the collective behaviour of particles emerging from baryon-rich fireballs created in heavy-ion interactions at this FAIR energy. These results are also useful for understanding of data measured at the RHIC Beam Energy Scan (BES) program.

\section{Acknowledgement}
\label{acknowledgement}
Sonia Kabana acknowledge the financial support received by ANID PIA/APOYO AFB220004. This research was supported in part by the cluster computing resource provided by the IT Division at the GSI Helmholtzzentrum für Schwerionenforschung, Darmstadt, Germany. The authors acknowledge helpful advices from the PHSD group members E. L. Bratkovskaya, V. Voronyuk, W. Cassing, P. Moreau, O. E. Soloveva, and L. Oliva. 

\section*{References}
\label{refs}

\end{document}